\documentstyle[12pt]{article}

\begin{document}


\title{QCD Corrections to  $b \to s \gamma$ Decay in 2-Higgs
Doublet Model}

\author{ Cai-Dian L\"{u}\thanks{ E-mail: lucd@itp.ac.cn.} \\
CCAST(World Laboratory), P.O.Box 8730, Beijing 100080, China\\
Institute of Theoretical Physics, Academia Sinica, P.O.Box 2735,\\
 Beijing 100080, China\thanks{Mailing address. }}

 \date{August 16, 1994}
\maketitle

\begin{abstract}
We give a more complete calculation of $b \to s\gamma $ decay
in 2-Higgs doublet model,
including leading log QCD corrections from $m_{top}$
to $M_W$ in addition to corrections  from $M_{W}$ to $m_b$.
The inclusive decay rate in the first model is found to be suppressed
23\% comparing with the calculations without the
QCD running from  $m_{top}$ to $M_W$. And the enhancement up to 5\% is found
in calculations of the second model. More strict restrictions to
parameters of 2-Higgs doublet model II are found.
\end{abstract}
\bigskip

\newpage

		\section{Introduction}

The standard model(SM)  has achieved a great success recent years.
However, there is still a vast interest beyond standard model.
One of the most simple extensions
of the standard model is to add another Higgs doublet. This may be
viewed as inevitable since many attempts at
going beyond the SM leads
to an enlargement of the Higgs sector (Such as Supersymmetry Model).
It is of interest to study the 2-Higgs doublet model, hence to constrain
other models.

The decay $b \to s \gamma$ is one of the very useful channel for study
of models beyond standard model\cite{Hew1}.
Recently the CLEO collaboration has observed\cite{cleo} the
exclusive decay $B \rightarrow K^* \gamma $ with a branching
 fraction
of $(4.5 \pm 1.5 \pm 0.9) \times 10^{-5}$. A new upper limit
on the inclusive $b \rightarrow s \gamma$ process is also
obtained as
$B(b \rightarrow s \gamma)<5.4 \times 10^{-4}$ at 95\%
C.L.\cite{cleo2}.
This leads to a number of papers\cite{Hew2} regarding this decay
recently. It has been argued that the experimental result is more close
to standard model predictions and provides more
information about restrictions on models beyond SM. This deeply depends
on more precise calculations of this decay.
The decay of $b\to s\gamma$ and its large leading log QCD corrections
have already been calculated in many papers\cite{Grin,Grig,Cel,Mis,Yao}.
Due to many reasons, there are discrepancies
among these papers, although there is not much differences in numerical
results.  Until recently, Ciuchini, Franco, Reina
and Silvestrini\cite{Ciu} performed this calculation in three regularization
schemes, to solve this problem completely. And also some efforts are
made to give a next to leading log calculations\cite{AJB,Ciu3}
 which is estimated
within 20\% contribution. All these efforts make it easy for
calculations of this decay in models beyond SM.

The radiative $b$ quark decay in 2-Higgs doublet model
has also been calculated in several  papers\cite{Grin,Wis,Hou,Ber}.
It is found to be strongly QCD-enhanced.
In other words, the strong interaction plays an important
role in this decay. However, there are still some uncertainties in
these  papers. All these papers do not include the
QCD running from $m_{top}$
to $M_W$. Since the top quark is found to be 2-times heavier than W boson(
$m_{top} = 174\pm 10^{+13}_{-12}$ GeV \cite{CDF} ),
and the charged Higgs is
also expected to be heavier(in supersymmetry model
 $m_{\phi}^2=M_W^2+m_A^2$),
it needs a detailed calculation of this effect.

In our present paper, by using effective field theory formalism,
we recalculate the $b \to s \gamma$ decay in 2-Higgs doublet model
including QCD running from $m_{top}$ to $M_W$, in addition to corrections
from $M_W$ to $m_b$, so as to give a complete leading log results.
In the next section, we first integrate out the top quark and charged Higgs,
generating an effective five-quark theory. By using the
renormalization group equation, we run the effective field theory
down to the W-scale.
In section 3, the weak bosons are removed. Then
we continue running the effective field theory
down to b-quark scale to include QCD corrections from $M_W$ to $m_b$.
In section 4, the rate of radiative $b$ decay is obtained.
Restrictions of model parameters from experiments
of CLEO collaboration are also given. Section 5 is a short summary.

 \section{QCD Corrections from $\mu =m_{top}$ to $\mu =M_W$ Scale}

With two Higgs doublets, one
has to avoid tree-level flavor changing neutral
currents (FCNCs) due to neutral Higgs bosons. This is usually achieved in
two ways, each involving some discrete symmetry. the first way (model I)
is to allow only one Higgs
doublet to couple to both types of quarks\cite{Hab}.
 The second way (model II) is to couple only one Higgs doublet to u-type
quarks while the other couples only to d-type quarks\cite{Gla}.
  It is of interest to note that model II
 occurs as a natural feature in theories with supersymmetry or a
Peccei-Quinn type of symmetry.

The Lagrangian of the 2-Higgs doublet  Model is
\begin{eqnarray}
 {\cal L}&= &
    \frac{1}{\sqrt{2}} \frac{\mu^{\epsilon/2}g_2}{M_W}
      \left[\left(\frac{v_2}{v_1}\right)
	\left(\begin{array}{ccc} \overline{u} & \overline{c} &
	      \overline{t} \end{array}\right)_R M_U V
	      \left(\begin{array}{c} d \\ s \\ b \end{array}\right)_L
	     -\xi
		\left(\begin{array}{ccc} \overline{u} & \overline{c} &
	     \overline{t} \end{array}\right)_L V M_D
	    \left(\begin{array}{c} d \\ s \\ b \end{array}\right)_R
	     \right]H_+ \nonumber\\
	&  +&h.c. \nonumber\\
	&  +&\cdot \cdot \cdot .
\end{eqnarray}
Where V represents the $3 \times 3$ unitary Kobayashi-Maskawa matrix,
$M_U$ and $M_D$ denote the diagonalized quark mass matrices, the
subscript
$L$ and $R$ denote left-handed and right-handed quarks, respectively.
For model I, $\xi={v_2}/{v_1}$; while for Model II, $\xi=-{v_1}/{v_2}$.
And $v_1$, $v_2$ are the magnitude of the vacuum expectation values of
two Higgs doublets, respectively. To keep explicit gauge invariance, we
 work in a background field  gauge\cite{Abbott}.

At first, we integrate out the top quark and charged Higgs, generating
an effective five quark theory, introducing dimension-6 effective
 operators as to include effects of the absent top quark and charged Higgs.
Higher dimension operators are suppressed by factor of $p^2/m_t^2$, where
$p^2$ characterizing the interesting external momentum of b quark
$p^2\sim m_b^2$. For leading order of $m_b^2/m_t^2$, dimension-6
operators are good enough to make a complete basis of operators:
\begin{eqnarray}
O_{LR}^1  & =  &  -\frac{1}{16\pi^2} m_b \overline{s}_L D^2 b_R,
\nonumber\\
O_{LR}^2  &  =  &  \mu^{\epsilon/2} \frac{g_3}{16\pi^2}
	m_b \overline{s}_L \sigma^{\mu\nu} X^a b_R G_{\mu\nu}^a,
\nonumber\\
O_{LR}^3  &  =  &  \mu^{\epsilon/2} \frac{e Q_b}{16\pi^2}
	m_b \overline{s}_L \sigma^{\mu\nu} b_R	F_{\mu\nu},
\nonumber\\
Q_{LR}  &  =  &  \mu^{\epsilon} g_3^2 m_b
	\phi_{+}\phi_{-} \overline{s}_L b_R,
\nonumber\\
P_L^{1,A}  &  =  &  -\frac{i}{16\pi^2}  \overline{s}_L
  T_{\mu\nu\sigma}^A D^{\mu} D^{\nu} D^{\sigma} b_L,\nonumber\\
P_L^2 & = & \mu^{\epsilon/2} \frac{e Q_b}{16\pi^2}  \overline{s}_L
	\gamma^{\mu} b_L \partial^{\nu} F_{\mu\nu},\nonumber\\
P_L^3  &  =  &  \mu^{\epsilon/2} \frac{e Q_b}{16\pi^2}  F_{\mu\nu}
	\overline{s}_L 	\gamma^{\mu} D^{\nu}b_L,\nonumber\\
P_L^4  &  =  &  i \mu^{\epsilon/2} \frac{e Q_b}{16\pi^2}
	\tilde{F}_{\mu\nu}
	\overline{s}_L \gamma^{\mu} \gamma^5 D^{\nu} b_L,\nonumber\\
R_L^1 &  =  &  i \mu^{\epsilon} g_3^2 \phi_{+}\phi_{-} \overline{s}_L
	\not \!\! D b_L,\nonumber\\
R_L^2  &  =  &  i \mu^{\epsilon} g_3^2(D^{\sigma} \phi_+) \phi_{-}
	\overline{s}_L\gamma_{\sigma} b_L,
\nonumber\\
R_L^3  &  =  &  i \mu^{\epsilon} g_3^2 \phi_{+}(D^{\sigma} \phi_{-})
	\overline{s}_L\gamma_{\sigma} b_L,\nonumber\\
W_{LR} &  =  & -i \mu^{\epsilon} g_3^2 m_b W^{\nu}_{+}W_{-}^{\mu}
	\overline{s}_L \sigma_{\mu \nu} b_R, \nonumber\\
W_L^1  &  =  &  i \mu^{\epsilon} g_3^2 W^{\nu}_{+}W_{-}^{\mu}
	\overline{s}_L
	\gamma _{\mu}\not \!\! D \gamma _{\nu} b_L,\nonumber\\
W_L^2  &  =  &  i \mu^{\epsilon} g_3^2(D^{\sigma} W^{\nu}_+) W^{\mu}_{-}
	\overline{s}_L \gamma_{\mu} \gamma_{\sigma} \gamma_{\nu} b_L,
\nonumber\\
W_L^3  &  =  &  i \mu^{\epsilon} g_3^2 W_{+\mu} W^{\mu}_{-}
\overline{s}_L \stackrel{\leftrightarrow}{\not \!\! D} b_L, \nonumber\\
W_L^4  &  =  &  i \mu^{\epsilon} g_3^2 W^{\nu}_+ W^{\mu}_{-}
\overline{s}_L (\stackrel{\leftrightarrow}{D}_{\mu}\! \gamma_{\nu} +
	 \gamma_{\mu}\! \stackrel{\leftrightarrow}{D}_{\nu} ) b_L.
\end{eqnarray}
Where $\overline{s}_L\! \stackrel{\leftrightarrow}{D}_{\mu}
\!\gamma_{\nu} b_L$
stands for $(\overline{s}_L D_{\mu} \gamma_{\nu} b_L +(D_{\mu}
 \overline{s}_L)
 \gamma_{\nu} b_L)$ and the covariant derivative is defined as
$$D_{\mu}=\partial_{\mu}-i\mu^{\epsilon/2}g_3 X^a G_{\mu}^{a} -
i \mu^{\epsilon/2}eQ A_{\mu},$$
with $g_3$ denoting the QCD coupling constant.
The tensor $T_{\mu\nu\sigma}^A$ appearing in $P_L^{1,A}$
 assumes the following
Lorentz structure, the index $A$ ranging from 1 to 4:
\begin{equation}
\begin{array}{ll}
        T_{\mu\nu\sigma}^1=g_{\mu\nu} \gamma_{\sigma},
&T_{\mu\nu\sigma}^2=g_{\mu\sigma} \gamma_{\nu},\nonumber\\
	T_{\mu\nu\sigma}^3=g_{\nu\sigma} \gamma_{\mu},
&T_{\mu\nu\sigma}^4=-i \epsilon_{\mu\nu\sigma\tau}
	\gamma^{\tau} \gamma_5.
\end{array}
\end{equation}
Then we can write down our effective Hamiltonian as
\begin{equation}
{\cal H}_{eff}=2 \sqrt{2} G_F V_{tb}V_{ts}^*\displaystyle \sum _i
C_i(\mu)O_i(\mu). \label{eff}
\end{equation}
The coefficients $C_i(\mu )$ can be calculated from
matching diagrams displayed in Fig.1 and Fig.2. Calculating
left hand side of Fig.1, keeping only leading orders of $p^2/m_t^2$,
we get coefficients of right hand side operators:
\begin{eqnarray}
C_{R_L^1} &=& C_{R_L^2}\;\;=\;\;-C_{Q_{LR}}\;\;=
	\;\;1/g_3^2,\nonumber\\
C_{R_L^3} &=& 0,\nonumber\\
C_{W_{LR}} &=& C_{W_L^3}\;\;=\;\;C_{W_{L}^4}\;\;=0,\nonumber\\
C_{W_L^1} &=& C_{W_L^2}\;\;=\;\;\delta /g_3^2.
\end{eqnarray}
In Fig.2, coefficients of right hand side operators
are all from the finite part integrations of left side
electroweak loops. Terms like $\log(\mu^2/m_t^2)$
vanish here, because of the matching scale $\mu=m_t$. They will
be regenerated by renormalization group running of electroweak later.
After calculation one has
\begin{eqnarray}
C_{O_{LR}^1}&=&-\left(\frac{1+\delta}
	{2(1-\delta)^2}+\frac{\delta}{(1-\delta)^3}\log\delta\right)
	-\xi'
	\left( \frac{ 1 +x }{2(1-x)^2}
	+\frac{x}{(1-x)^3}\log x \right), \nonumber\\
C_{O_{LR}^2}&=&-\frac{1}{2} \left(\frac{1}{(1-\delta)}+
	\frac{\delta}{(1-\delta)^2}\log\delta\right)
  	-\xi'
	\left( \frac{ 1 }{2(1-x)}
	+\frac{x}{2(1-x)^2}\log x \right), \nonumber\\
C_{O_{LR}^3}&=& \left(\frac{1}{(1-\delta)}+
	\frac{\delta}{(1-\delta)^2}\log\delta\right)
	+\xi
	\left( \frac{ 1 }{1-x} +\frac{x}{(1-x)^2}\log x \right), \nonumber\\
C_{P_L^{1,1}}&=& C_{P_L^{1,3}}\;\;=\;\;
	\left(\frac{\frac{11}{18}+\frac{5}{6}\delta
	-\frac{2}{3}\delta^2 +\frac{2}{9} \delta ^3}{(1-\delta)^3}+
\frac{\delta+\delta^2-\frac{5}{3}\delta^3 +\frac{2}{3} \delta ^4}
	{(1-\delta)^4}\log\delta\right)\nonumber\\
  &&~~~~~~~~+\left(\frac{v_2}{v_1}\right)^2 \left(
	\frac{\frac{11}{18} -\frac{7}{18}x +\frac{1}{9}x^2 }{ (1-x)^3}
	+\frac{x-x^2+\frac{1}{3}x^3 }{ (1-x)^4 }\log x \right),\nonumber\\
C_{P_L^{1,2}}&=&\left(\frac{-\frac{8}{9}-\frac{1}{6}\delta
	+\frac{17}{6}\delta^2 -\frac{7}{9} \delta ^3}{(1-\delta)^3}+
	\frac{-\delta+\frac{10}{3}\delta^3 -\frac{4}{3} \delta^4}
	{(1-\delta)^4}\log \delta \right)\nonumber\\
  &&+\left(\frac{v_2}{v_1}\right)^2 \left(
	\frac{-\frac{8}{9} +\frac{29}{18}x -\frac{7}{18}x^2 }{ (1-x)^3}
	+\frac{-x+2x^2-\frac{2}{3}x^3 }{ (1-x)^4 }\log x \right),\nonumber\\
C_{P_L^{1,4}}&=&\left(\frac{\frac{1}{2}-\delta
	-\frac{1}{2}\delta^2 +\delta^3}{(1-\delta)^3}+
\frac{\delta-3\delta^2+2\delta^3}{(1-\delta)^4}\log\delta\right)\nonumber\\
  &&+\left(\frac{v_2}{v_1}\right)^2 \left(
	\frac{1 -x^2 }{ 2(1-x)^3}
	+\frac{x-x^2 }{ (1-x)^4 }\log x \right),\nonumber\\
C_{P_L^2}&=&\frac{1}{Q_b}\left(\frac{\frac{3}{4}+\frac{1}{2}\delta
	-\frac{7}{4}\delta^2 +\frac{1}{2} \delta^3 }{(1-\delta)^3}
	-\frac{1}{3} \delta
+\left(\frac{\frac{1}{6} +\frac{5}{6}\delta -\frac{5}{3}\delta^3+
	\frac{2}{3} \delta^4} {(1-\delta)^4}
	- \frac{1}{6} -\frac{1}{3} \delta \right)
	\log\delta \right)\nonumber\\
  &&+\frac{1}{Q_b}\left(\frac{v_2}{v_1}\right)^2 \left(
	\frac{\frac{3}{4} -x +\frac{1}{4}x^2 }{ (1-x)^3}
	+\frac{ \frac{1}{6} +\frac{1}{2}x -x^2+\frac{1}{3}x^3 }
	{ (1-x)^4 }\log x \right),\nonumber\\
C_{P_L^3}&=&0,\nonumber\\
C_{P_L^4}&=&\frac{1}{Q_b}\left(\frac{-\frac{1}{2}-5\delta
	+\frac{17}{2}\delta^2 -3\delta^3 }{(1-\delta)^3}+
	\frac{-5\delta +7\delta^2 -2\delta^3}{(1-\delta)^4}\log\delta
	+4\delta \log\delta \right)\nonumber\\
  &&-\frac{1}{Q_b}\left(\frac{v_2}{v_1}\right)^2 \left(
	\frac{1 -x^2 }{ 2(1-x)^3}
	+\frac{x-x^2 }{ (1-x)^4 }\log x \right).\label{coe}
\end{eqnarray}
Where $\delta = M_W^2/m_t^2$, $x=m_{\phi}^2 /m_t^2$;
With
$$\xi'=v_2^2/v_1^2,~~~~~ model~ I, $$
$$\xi'=-1,~~~~~~~  model~ II.$$
When $\xi =\xi'=0$, the above result (\ref{coe}) reduces to that of SM
 case\cite{Cho,lcd1}.

The renormalization group equation satisfied by
the coefficient functions $C_i(\mu)$ is
\begin{equation}
\mu \frac{d}{d\mu} C_i(\mu)=\displaystyle\sum_{j}(\gamma^{\tau})_{
ij}C_j(\mu).\label{ren}
\end{equation}
Where the anomalous dimension matrix $\gamma_{ij}$
is calculated in practice by requiring
renormalization group equations for Green functions
with insertions of composite operators
to be satisfied order by order in perturbation theory.

After evaluating the loop diagrams, we find the following
leading order weak mixing of operators, with the Q, R part
 agrees with ref.\cite{Cho}.
\begin{equation}
\begin{array}{rccl}
  \gamma=
   & \begin{array}{c} \\ Q_{LR}\\ R_L^1\\ R_L^2\\ R_L^3\\ W_{LR}\\
	 W_L^1\\	W_L^2\\ W_L^3\\ W_L^4  \end{array}
   & \begin{array}{c}
  	\begin{array}{cccccccc}
O_{LR}^1& O_{LR}^2& O_{LR}^3 & P_L^{1,A} & P_L^2 & P_L^3 & P_L^4 &
  	\end{array} \\

	\left(\begin{array}{ccccccccccccc}
	 0 && 0 &&& 0 && 0 &  0   & 0 && 0 \\
	 0 && 0 &&& 0 && 0 &  0   & 0 && 0 \\
         0 && 0 &&& 0 && 0 & -1/2 & 0 && 0 \\
         0 && 0 &&& 0 && 0 & 1/2  & 0 && 0 \\
	 0 && 0 &&& 6 && 0 &  0   & 0 && 0\\
	 0 && 0 &&& 0 && 0 &  0   & 0 && 12\\
	 0 && 0 &&& 0 && 0 & -1   & 0 && 0 \\
	 0 && 0 &&& 0 && 0 &  0   & 0 && 0 \\
	 0 && 0 &&& 0 && 0 &  0   & 0 && 0 \\
	\end{array}\right)

     \end{array}

   & 16\pi^2\; \displaystyle{ \frac{g_3^2}{8\pi^2} }.

\end{array}\label{weak}
\end{equation}
These mixing are all between operators induced by tree-diagrams and
operators induced by loop-diagrams.
The vanishing $\log(\mu^2/m_t^2)$ terms in previous matching
are regenerated here by renormalization group equation(\ref{ren}).

The QCD anomalous dimensions for each of the operators
in our basis are
\begin{equation}
 \begin{array}{c}
     \begin{array}{ccccccccccccc}
	& && O_{LR}^1 & O_{LR}^2& O_{LR}^3& P_L^{1,1}& P_L^{1,2}&
	  P_L^{1,3}& P_L^{1,4}& P_L^{2}& P_L^{3}& P_L^{4}
     \end{array}\\
     \begin{array}{r}
  O_{LR}^1\\ O_{LR}^2\\ O_{LR}^3\\ P_L^{1,1}\\ \gamma=\; P_L^{1,2}\\
	  P_L^{1,3}\\ P_L^{1,4}\\ P_L^{2}\\ P_L^{3}\\ P_L^{4}
     \end{array}\left(\begin{array}{ccccccccccccccc}
	  \frac{20}{3} && 1 && -2 & 0 & 0 & 0 & 0 && 0 & 0 && 0 \\
 -8 && \frac{2}{3} && \frac{4}{3} & 0 & 0 & 0 & 0 && 0 & 0 && 0 \\
	 0 && 0 && \frac{16}{3} & 0 & 0 & 0 & 0 && 0 & 0 && 0 \\
	 6 && 2 && -1 & \frac{2}{3} & 2 & -2 & -2 && 0 & 0 && 0 \\
	 4 && \frac{3}{2} && 0 & -\frac{113}{36} & \frac{137}{18}
	 & -\frac{113}{36} &-\frac{4}{3} &&\frac{9}{4} & 0 && 0  \\
	 2 && 1 && 1 & -2 & 2 & \frac{2}{3} & -2 && 0 & 0 && 0 \\
	 0 && \frac{1}{2} && 2 & -\frac{113}{36} & \frac{89}{18}
	  & -\frac{113}{36} &
	 \frac{4}{3} && \frac{9}{4} & 0 && 0 \\
	 0 && 0 && 0 & 0 & 0 & 0 & 0 && 0 & 0 && 0 \\
	 0 && 0 && -\frac{4}{3} & 0 & 0 & 0 & 0 && 0 & 0 && 0 \\
	 0 && 0 && -\frac{4}{3} & 0 & 0 & 0 & 0 && 0 & 0 && 0
	\end{array}\right) \displaystyle{ \frac{g_3^2}{8\pi^2} },
\end{array}
\label{anom1}
\end{equation}

\begin{equation}
\begin{array}{cccc}
\gamma= & \begin{array}{c}
	\\ Q_{LR}\\ R_L^1\\ R_L^2\\ R_L^3\\ W_{LR}\\ W_L^1\\ W_L^2\\
		W_L^3\\ W_L^4\\
          \end{array}

	& \begin{array}{c}
	      \begin{array}{ccccccccc}Q_{LR} & R_L^1 & R_L^2 & R_L^3
			  & W_{LR} & W_L^1 & W_L^2 & W_L^3 & W_L^4
 		     \end{array}
	    \\
	      \left( \begin{array}{ccccccccccccc}
 \frac{23}{3} && 0  & 0  && 0 & 0 && 0 && 0 &  0  &      0\\
 0  && \frac{23}{3} & 0  && 0 & 0 && 0 && 0 &  0  &     0\\
 0  &&  0 & \frac{23}{3} && 0 & 0 && 0 && 0 &  0  &    0\\
 0  &&  0 & 0 && \frac{23}{3} & 0 && 0 && 0 &  0  &     0\\
 0  &&  0   &   0   &&  0    & 13 && 0 && 0  &  0  &     0\\
 0 && 0& 0 && 0 &-\frac{8}{3} &&\frac{23}{3} && 0 &-\frac{8}{9}
&\frac{16}{9}\\
 0  &&  0 & 0 && 0 & 0 && 0 && \frac{23}{3} &  0  &    0\\
 0  &&  0 & 0 && 0 & 0 && 0 && 0  & \frac{23}{3} &    0\\
 0  &&  0 & 0 && 0 & 0 && 0 && 0 & -\frac{16}{9} & \frac{101}{9}\\
	      \end{array} \right)
	  \end{array}
	& \displaystyle{ \frac{g_3^2}{8 \pi^2} }.

\end{array}\label{anom2}
\end{equation}

The mixing elements between the two matrix (\ref{anom1}) and
(\ref{anom2}) are all zero in leading log approximation.
There are some differences in the anomalous
dimension matrix comparing with Cho and Grinstein's result\cite{Cho}.
If we omit a symmetric factor of 1/2 in calculating Feynman diagram like
Fig.3, the results will agree with them\cite{lcd1}.
But according to similar diagram
calculation like ref.\cite{Yao,Ciu}, this factor can not be omitted.
For instance, diagram like Fig. 3 is also appeared
in calculation of anomalous dimension of the gluon magnetic moment-type
operator $O_8$ in ref.\cite{Yao,Ciu} with a symmetric factor 1/2.
This is also shown in general Feynman gauge calculation, and it give results
same as the above. After these changes, the whole matrix can be easily
diagonalized, and gives all real eigenvalues, which is required to maintain
hermiticity of the effective hamiltonian at all renormalization scales.
While in ref.\cite{Cho}, it can not.
In their case, some eigenvalues are complex.

The solution to renormalization group equation (\ref{ren})
appears in obvious matrix notation as
\begin{equation}
C(\mu_2)=\left[\exp\int_{g_3(\mu_1)}^{g_3(\mu_2)}dg\frac
{\gamma^T(g)}{\beta(g)}\right] C(\mu_1).\label{solu}
\end{equation}
After inserting anomalous dimension (\ref{weak}--\ref{anom2}),
we can have the coefficients of operators at $\mu=M_W$.

\section{QCD Corrections from $\mu=M_W$ to $\mu =m_b$ Scale}

In order to continue running the basis operator coefficients down to
lower scales, one must integrate out the weak gauge bosons and
would-be
Goldstone bosons at $\mu=M_W$ scale. The diagrams are displayed in
Fig.4. From  the second and third matching equations of this figure,
one finds the following relations
between coefficient functions just below(-) and above(+) $\mu=M_W$:
\begin{eqnarray}
C_{O_{LR}^1}(M_W^-)&=&C_{O_{LR}^1}(M_W^+), \nonumber\\
C_{O_{LR}^2}(M_W^-)&=&C_{O_{LR}^2}(M_W^+), \nonumber\\
C_{O_{LR}^3}(M_W^-)&=&C_{O_{LR}^3}(M_W^+),\nonumber \\
C_{P_L^{1,1}}(M_W^-)&=&C_{P_L^{1,1}}(M_W^+) + 2/9,\nonumber \\
C_{P_L^{1,2}}(M_W^-)&=&C_{P_L^{1,2}}(M_W^+) - 7/9, \nonumber\\
C_{P_L^{1,3}}(M_W^-)&=&C_{P_L^{1,3}}(M_W^+) + 2/9,\nonumber \\
C_{P_L^{1,4}}(M_W^-)&=&C_{P_L^{1,4}}(M_W^+) + 1, \nonumber\\
C_{P_L^2}(M_W^-)&=&C_{P_L^2}(M_W^+)
	-C_{W_L^2}(M_W^+) - 3/2, \nonumber\\
C_{P_L^3}(M_W^-)&=&C_{P_L^3}(M_W^+), \nonumber\\
C_{P_L^4}(M_W^-)&=&C_{P_L^4}(M_W^+) + 9.
\end{eqnarray}
In addition to these, there are new four-quark operators from
the first equation of Fig.4\cite{Grin,Mis,Ciu}:
\begin{eqnarray}
O_1&=&(\overline{c}_{L\beta} \gamma^{\mu} b_{L\alpha})
	    (\overline{s}_{L\alpha} \gamma_{\mu} c_{L\beta}),
\nonumber\\
O_2&=&(\overline{c}_{L\alpha} \gamma^{\mu} b_{L\alpha})
	    (\overline{s}_{L\beta} \gamma_{\mu} c_{L\beta}),
\nonumber\\
O_3&=&(\overline{s}_{L\alpha} \gamma^{\mu} b_{L\alpha})
	    [(\overline{u}_{L\beta} \gamma_{\mu} u_{L\beta})+...+
	    (\overline{b}_{L\beta} \gamma_{\mu} b_{L\beta})],
\nonumber\\
O_4&=&(\overline{s}_{L\alpha} \gamma^{\mu} b_{L\beta})
	    [(\overline{u}_{L\beta} \gamma_{\mu} u_{L\alpha})+...+
	    (\overline{b}_{L\beta} \gamma_{\mu} b_{L\alpha})],
\nonumber\\
O_5&=&(\overline{s}_{L\alpha} \gamma^{\mu} b_{L\alpha})
	    [(\overline{u}_{R\beta} \gamma_{\mu} u_{R\beta})+...+
	    (\overline{b}_{R\beta} \gamma_{\mu} b_{R\beta})],
\nonumber\\
O_6&=&(\overline{s}_{L\alpha} \gamma^{\mu} b_{L\beta})
	    [(\overline{u}_{R\beta} \gamma_{\mu} u_{R\alpha})+...+
	    (\overline{b}_{R\beta} \gamma_{\mu} b_{R\alpha})],
\end{eqnarray}
with coefficients
$$ C_i(M_W)=0, \;\; i=1,3,4,5,6, \;\; C_2(M_W)=1.$$

     To simplify the calculation and compare with the previous results,
equations of motion(EOM)\cite{eom} is used to reduce all the remaining
two-quark operators to the gluon and photon magnetic moment
operators $O_{LR}^2$ and $O_{LR}^3$.
To be comparable with previous results without QCD corrections from $m_{top}$
to $M_W$, operators $O_{LR}^3$, $O_{LR}^2$ are rewritten
as $O_7$, $O_8$ like ref.\cite{Grin,Mis,Ciu},
\begin{eqnarray}
O_7&=&(e/16\pi^2) m_b \overline{s}_L \sigma^{\mu\nu}
	    b_{R} F_{\mu\nu},\nonumber\\
O_8&=&(g/16\pi^2) m_b \overline{s}_{L} \sigma^{\mu\nu}
	    T^a b_{R} G_{\mu\nu}^a.
\end{eqnarray}
Then
\begin{eqnarray}
C_7(M_W^-) &=& \frac{1}{3} C_{O_{LR}^3}(M_W^-), \nonumber\\
C_8(M_W^-) &=& - C_{O_{LR}^2}(M_W^-).
\end{eqnarray}

The operator basis now consists of 8 operators.
The  effective Hamiltonian appears just below the W-scale as
\begin{eqnarray}
{\cal H}_{eff} && =\frac{4G_F}{\sqrt{2}} V_{tb}V_{ts}^*
	\displaystyle \sum_{i}
                C_i(M_W^-) O_i(M_W^-)\nonumber\\
	& &\stackrel{EOM}{\rightarrow}
		\frac{4G_F}{\sqrt{2}} V_{tb}V_{ts}^*\left\{
		\displaystyle{\sum_{i=1}^{6} }C_i (M_W^-)O_i + C_7(M_W^-) O_7
		+C_8(M_W^-) O_8 \right\}.
\end{eqnarray}

For completeness, the explicit expressions of
the coefficient of operator $O_8$ and $O_7$ at $\mu=M_W^-$ are given,
\nopagebreak[1]
\begin{eqnarray}
C_{O_8}(M_W^-) &= & \left( \frac{\alpha _s (m_t)} {\alpha _s (M_W)}
	\right) ^{ \frac{14}{23} } \left\{ \frac{1}{2}C_{O_{LR}^1}(m_t)
-C_{O_{LR}^2}(m_t) +\frac{1}{2}C_{P_L^{1,1}}(m_t) \right.\nonumber\\
&&	\;\;\;\;\;\;\;\;\;\;\;\;\;\;\;\;\;\;
	\left.+\frac{1}{4}C_{P_L^{1,2}}(m_t)
	-\frac{1}{4}C_{P_L^{1,4}}(m_t)\right\}
	-\frac{1}{3} ,\label{c2}
\end{eqnarray}
\begin{eqnarray}
&\displaystyle{
C_{O_7}(M_W^-) = \frac{1}{3}\left( \frac{\alpha _s (m_t)} {\alpha _s (M_W)}
	\right) ^{ \frac{16}{23} } \left\{ C_{O_{LR}^3}(m_t)
	+8 C_{O_{LR}^2}(m_t) \left[1-\left( \frac{\alpha _s (M_W)}
{\alpha _s (m_t)} \right) ^{ \frac{2}{23} } \right]\right.}&\nonumber\\
&\displaystyle{	+\left[-\frac{9}{2} C_{O_{LR}^1}(m_t)
	-\frac{9}{2}C_{P_L^{1,1}}(m_t)
-\frac{9}{4}C_{P_L^{1,2}}(m_t) +\frac{9}{4}C_{P_L^{1,4}}(m_t)\right]
\left[1- \frac{8}{9} \left( \frac{\alpha _s (M_W)}
{\alpha _s (m_t)} \right) ^{ \frac{2}{23} } \right] }&\nonumber\\
&\displaystyle{	\;\;\;\left.-\frac{1}{4}C_{P_L^4}(m_t)
+\frac{9}{23} 16\pi^2 C_{W_L^1}(m_t) \left[1- \frac{\alpha _s (m_t)}
	{\alpha _s (M_W)} \right] \right \}
	-\frac{23}{36} }. &  \label{c3}
\end{eqnarray}
Since they are expressed by coefficients of operators
at $\mu=m_t$ and QCD coupling $\alpha_s$,
it is convenient to utilize these formula.

If the QCD corrections from $m_{top}$ to $M_W$ are ignored (by
setting $\alpha_s(m_t)=\alpha_s(M_W)$
 in eqn.(\ref{c2}),(\ref{c3}) ), the above results(\ref{c2})(\ref{c3})
 would reduce to the previous results\cite{Grin,Ber} exactly,
where the top quark and W bosons are integrated out together:
\begin{eqnarray}
C_7(M_W)&=& -\frac{1}{2} A(x)-\frac{1}{6} \left(\frac{v_2}{v_1}\right)^2
	A(y) +\xi' B(y)\\
C_8(M_W)&=& -\frac{1}{2} D(x)-\frac{1}{6} \left(\frac{v_2}{v_1}\right)^2
	D(y) +\xi' E(y),
\end{eqnarray}
with A(x), B(y), D(x), E(y) defined in ref.\cite{Grin}.

The effects of QCD corrections to $C_7(M_W)$ and
$C_8(M_W)$ can easily be seen from Fig.5 and Fig.6.
Here W boson mass is taken as $M_W=80.22$GeV, The top quark mass $m_t=174$GeV,
and the QCD scale is taken as $\Lambda_{QCD}^{f=5}=175$ MeV\cite{data},
$M_{H^{+-}}=300GeV$.
The results of model I is displayed in Fig.5. Except for small
values of $v_2/v_1$, $C_7(M_W)$ and $C_8(M_W)$ are both suppressed
by QCD corrections from $m_t$ to $M_W$. At $v_2/v_1=10$, $C_7(M_W)$ is
suppressed 17\% and $C_8(M_W)$ 12\%.
Fig.6 gives results calculated in model II, the
differences are that $|C_7(M_W)|$ and $|C_8(M_W)|$ are both enhanced
by QCD corrections for all values of $v_2/v_1$. At small values of $v_2/v_1$,
$|C_7(M_W)|$ is enhanced 10\%, $|C_8(M_W)|$ is enhanced 8\%.
Since $C_7(M_W)$ and $C_8(M_W)$ are both the input of the following QCD
running from $M_W$ to $m_b$,
It is expected to change the final result.

The running of the coefficients of operators from $\mu=M_W$ to $\mu=m_b$
was well described in ref.\cite{Mis,Ciu}. After this running we have
the coefficients of operators at $\mu=m_b$ scale.
Here $m_b=4.9$GeV is used. Except small values of $v_2/v_1$,
both $C_7(m_b)$ and $C_8(m_b)$ calculated in model I are suppressed
in comparison to values obtained by
ref.\cite{Grin,Ciu2}, where the QCD running from $m_t$ to $M_W$ is
neglected. While in model II both $|C_7(m_b)|$ and $|C_8(m_b)|$ are enhanced.

\section{The $\overline{B} \rightarrow X_s \gamma$ decay rate}

The leading order $b \rightarrow s\gamma$ matrix element of $H_{eff}$
is given by the sum of operators $O_5$, $O_6$ and $O_7$ in our
effective theory\cite{Mis,Ciu},
\begin{equation}
<H_{eff}>=-2 \sqrt{2} G_F V_{ts}^* V_{tb} \left\{ C_7(\mu)+Q_d
[C_5(\mu)+3 C_6
(\mu)] \right\} <|O_7|>. \label{Heff}
\end{equation}
Therefore, the sought amplitude will be proportional to the squared
 modulus of
\begin{equation}
	C_7^{eff}(m_b)=C_7(m_b)+Q_d\;[C_5(m_b)+3 C_6(m_b)] \label{C7}
\end{equation}
instead of $|C_7(m_b)|^2$ itself.

Following ref.\cite{Grin,Mis,Ciu},
\begin{equation}
BR(\overline{B} \rightarrow X_s \gamma) /BR(\overline{B}
\rightarrow X_c e\overline{\nu}) \simeq\Gamma(b\rightarrow
s\gamma)/\Gamma
(b\rightarrow ce\overline{\nu}).
\end{equation}
Then applying eqs.(\ref{Heff}),(\ref{C7}), one finds
\begin{equation}
\frac{BR(\overline{B} \rightarrow X_s \gamma)}{BR(\overline{B}
\rightarrow X_c e \overline{\nu})} \simeq \frac{6 \alpha_{QED}}{\pi
 g (m_c/m_b)}
|C_7^{eff}(m_b)|^2 \left(1-\frac{2 \alpha_{s}(m_b)}{3 \pi} f(m_c/m_b)
\right)^
{-1},
\end{equation}
where $g(m_c/m_b)\simeq 0.45$ and $f(m_c/m_b)\simeq 2.4$ corresponding
to the phase space
factor and the one-loop QCD correction to the semileptonic decay,
respectively\cite{Cabi}. The electromagnetic fine structure constant
evaluated at the $b$ quark scale takes value as $\alpha_{QED}(m_b)=
1/132.7$.  Afterwards one obtains the $\overline{B} \rightarrow X_s
\gamma$ decay rate normalized to the quite well established
 semileptonic decay rate.
If we take experimental result $Br(\overline{B} \to
X_c e\overline{\nu} ) =10.8\% $\cite{data}, the branching ratios of
$\overline{B} \to X_s \gamma$ is found.

The decay results are summarized in Fig.7 and Fig.8 as functions of
$v_2/v_1$, with different charged Higgs mass 170GeV, 300GeV, 600GeV
and 900GeV. The CLEO upper limit is also shown as a solid line.

For 2-Higgs doublet model I, the decay rates including QCD running
from $m_t$ to $M_W$ are suppressed except small values of $v_2/v_1$.
At $v_2/v_1=10$, $m_{H^{+-}}=300GeV$, the suppression is 23\%. As
$ v_2/v_1 \to 0$, this kind of model goes back to SM. In this small
$ v_2/v_1$ region, the QCD corrections give 7\% enhancement which
corresponding to SM case\cite{lcd1}.
Therefore, the restrictions to this model parameters are less
tight than previous predictions especially for large values of
$ v_2/v_1$. In Fig.7, one can see that, for lower
 mass of charged Higgs, small values of $v_2/v_1$ are still  allowed.
The parameter space is still open. The decay channel $t \to b \phi$
can still exist for a wide region.

The decay rates in 2-Higgs doublet model II are enhanced up to 5\% more than
previous calculations when charged Higgs mass is lower than 800 GeV.
 Although this percentage is not very large, the absolute
values of the decay rates are sure higher. In Fig.8, one can easily see that,
as charged Higgs mass lower than 700GeV, all values of $v_2/v_1$ are
excluded by CLEO's $b \to s \gamma$ experiments.
For larger charged Higgs mass, small value of $v_2/v_1$ is
still allowed. It is obvious that, the top decay channel $t \to b \phi$
is already ruled out in this kind of 2-Higgs doublet model.

In the Supersymmetry model, the Higgs sector is the same as model II, but
the large chargino contribution cancels much the charged Higgs
contribution\cite{Barb}.
There is sure a suppression of decay rates from 2-Higgs doublet model II.
The magnitude is more complicated, since it depends on various Supersymmetry
models and chargino mass. When chargino mass is of order O($m_{top}$)
or higher, it should be integrated out at $\mu=m_{top}$; if it is of order
O($M_W$) or lower, it should be integrated out at $\mu=M_W$ scale. Further
more, for a complete Supersymmetry calculation, there are also contributions
from other SuSy particles which need a detailed calculation\cite{lcd2}.

\section{Conclusion}

As a conclusion, we have given the full leading log QCD
corrections(including
QCD running from $m_{top}$ to $M_W$) to $b\to s \gamma$ decay
in 2-kind 2-Higgs doublet models.

The QCD running from $m_t$ to $M_W$ suppresses the $b\to s \gamma$
decay rate in model I, and enhances the decay rate in model II.

Although this result is not quite different from
the previous calculations, our improvements lie in reducing some
theoretical uncertainties. After these changes,  restrictions from
$b\to s\gamma$ decay to
2-Higgs doublet model parameters are less strict in model I and
more tight in model II than previous predictions.
It is shown that the decay $b\to s\gamma$ is by far the most
restrictive process in constraining the parameters of the charged
Higgs boson sector in 2-Higgs doublet model.

\noindent
{\it Note added:}

After this paper was submitted for publication, the
paper\cite{Anl} by H. Anlauf came to our attention; the author studied
the supersymmetric contributions as well as charged Higgs contributions
in 2HD model II. Our results consist of 2-kinds of 2HD models; and
agree with their result of model II at leading order($M_W/m_t$, $m_H/m_t$).
Further more, we also included high order contribution which can not
be neglected(In ref.\cite{Anl}, this was picked up later after running
to $M_W$ scale to match onto results without QCD running from $m_t$
to $M_W$).

\section*{{Acknowledgement}}

The author thanks Prof. X.Y. Li, Z.M. Qiu,
Z.X. Zhang, and Dr. Y.Q. Chen, Q.H. Zhang for
helpful discussions.

\section*{Figure Captions}
\parindent=0pt

Fig.1 Leading order matching conditions at the top quark scale for the
1PI Green functions  in the full theory and in the intermediate
effective field theory.

Fig.2 One loop matching conditions at the top quark scale for the
1PI Green functions  in the full theory and in the intermediate
effective field theory.

Fig.3 One of the Feynman diagram in calculating anomalous dimensions,
with the heavy dot denoting high dimension operator.

Fig.4 Matching conditions at $\mu=M_W$ for four quarks and two
quarks 1PI Green functions in the intermediate
effective field theory and effective field theory below W scale.

Fig.5 The photon and gluon magnetic moment operator's coefficients
$C_7(M_W)$ and $C_8(M_W)$ in model I. The solid lines are our QCD corrected
results, the other two are uncorrected ones.

Fig.6 The photon and gluon magnetic moment operator's coefficients
$C_7(M_W)$ and $C_8(M_W)$ in model II. The solid lines are our QCD corrected
results, the other two are uncorrected ones.

Fig.7 BR($\overline{B} \rightarrow X_s \gamma$) of model I
	as function of $v_2/v_1$ for different charged Higgs masses.
The solid line is the upper limit of CLEO. This line upper is the
excluded region.

Fig.8 BR($\overline{B} \rightarrow X_s \gamma$) in model II
	as function of $v_2/v_1$ for different charged Higgs masses.
 The solid line is the upper limit of CLEO.

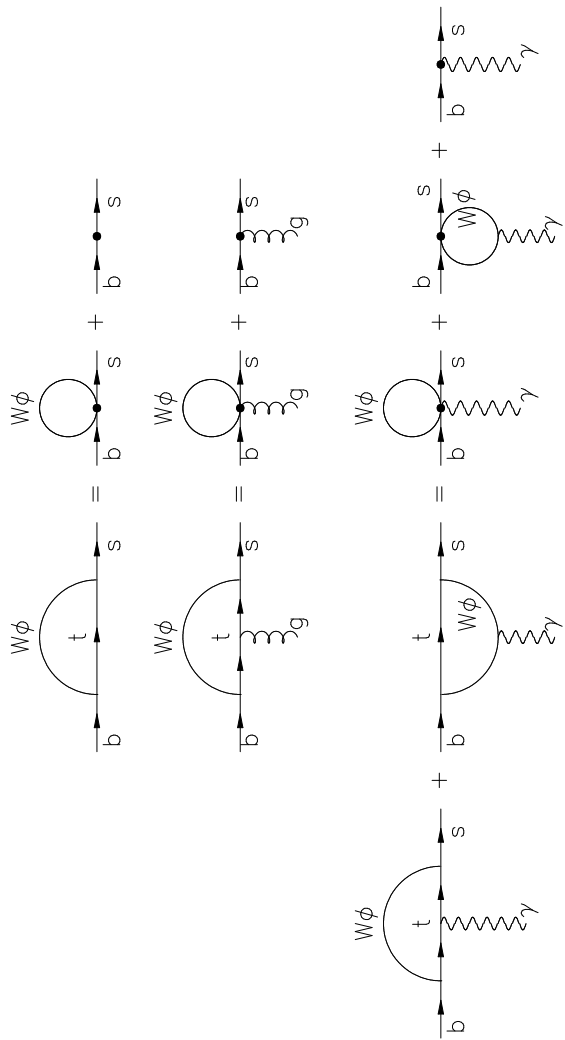
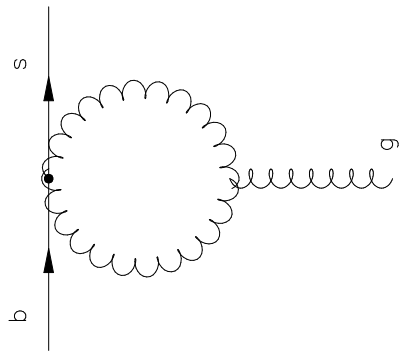
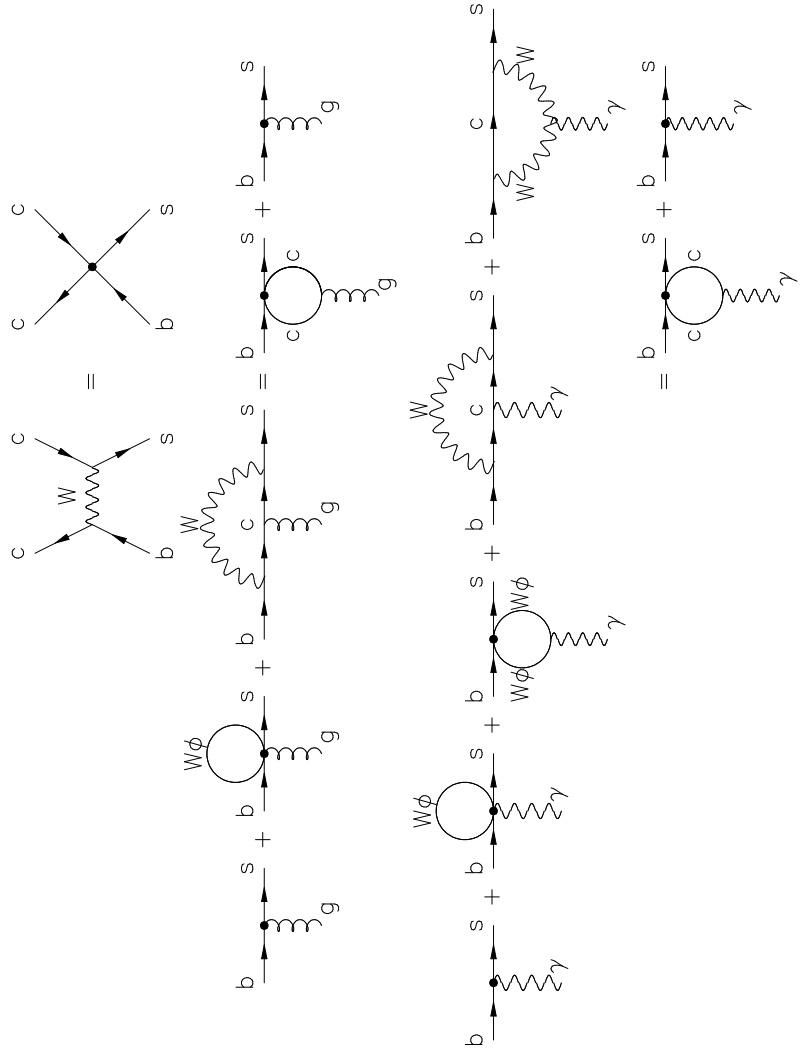
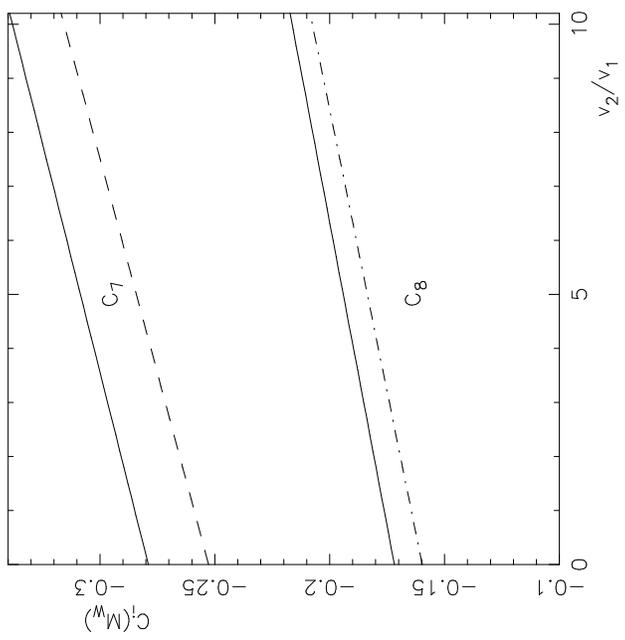
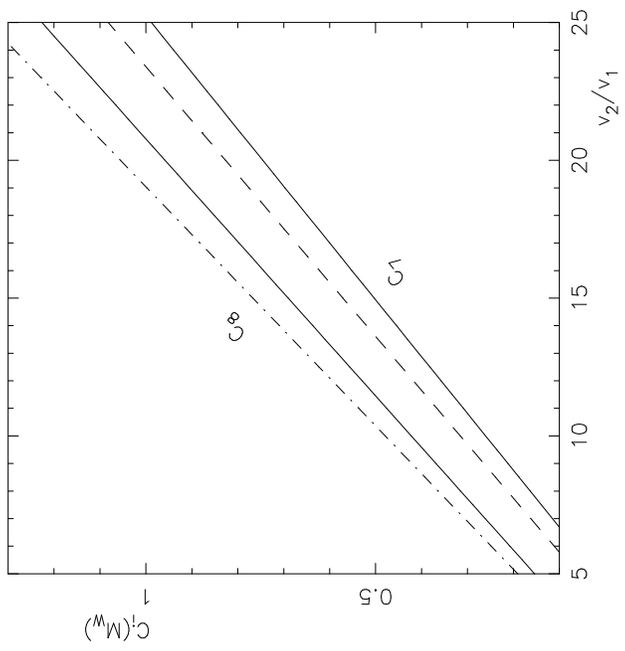
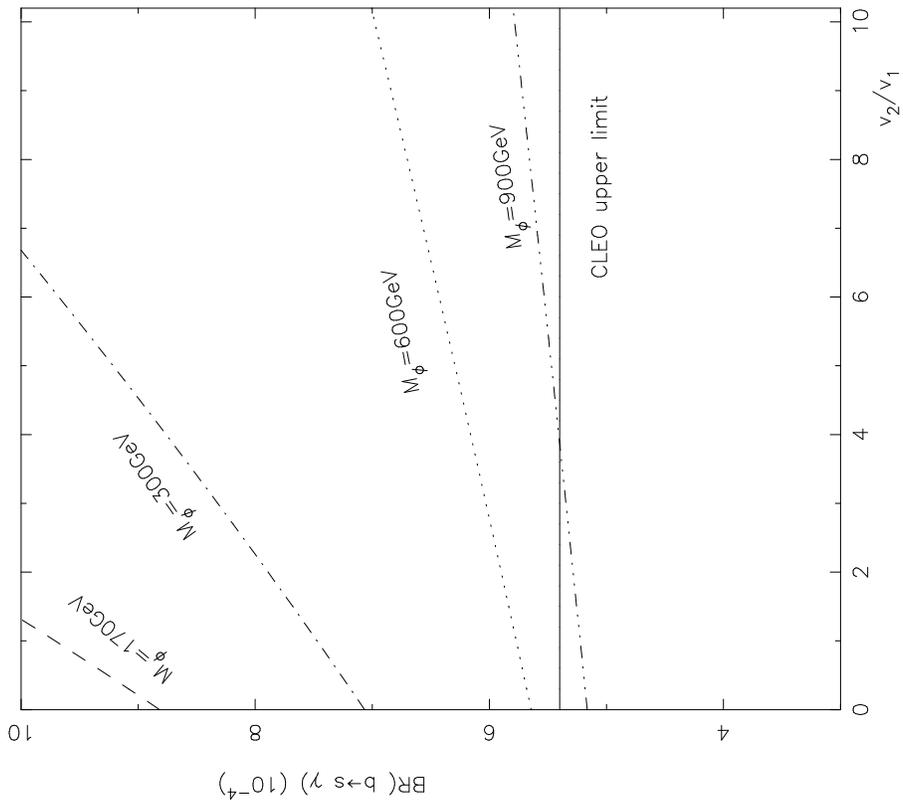
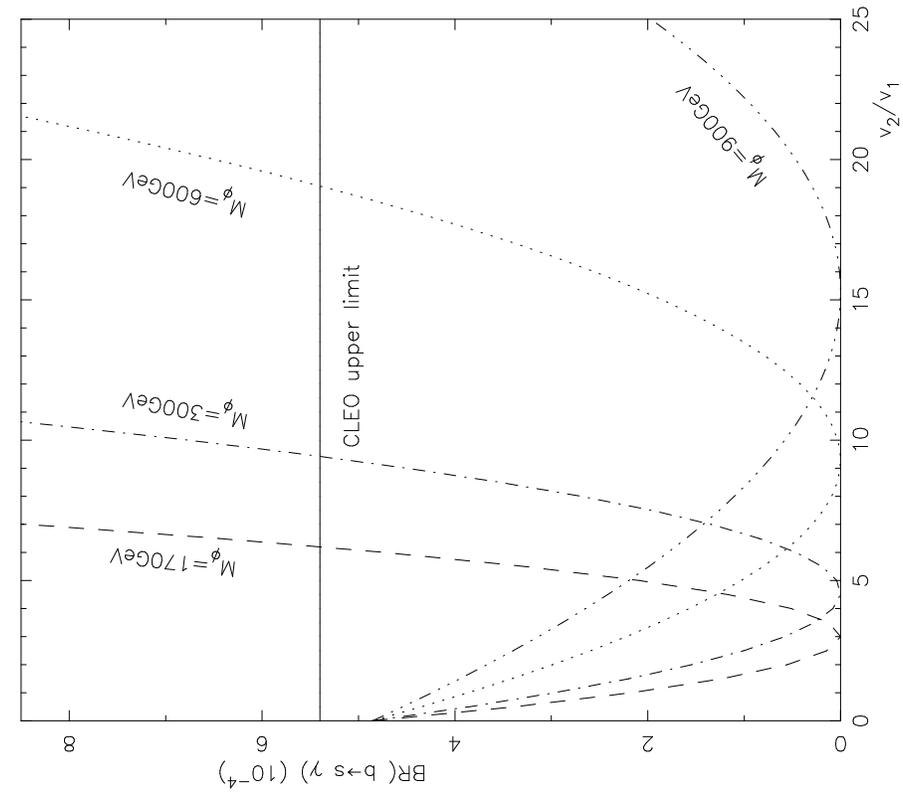
\end{document}